\documentclass[aps,prl,twocolumn,showpacs,amsmath,amssymb,floatfix,longbibliography,footinbib,superscriptaddress,10pt]{revtex4-2}
\usepackage{graphicx,mathtools,bm,array,enumitem,overpic,booktabs,multirow}
\usepackage{natbib}
\usepackage[dvipsnames]{xcolor} 
\usepackage{braket}
\usepackage{float}
\usepackage{textcomp}
\usepackage{comment}
\usepackage[normalem]{ulem}
\usepackage[bookmarks=true,colorlinks,linkcolor=RoyalBlue,urlcolor=RoyalBlue,citecolor=RoyalBlue]{hyperref}

\newcommand{\sA}{\mathcal{A}}
\newcommand{\sB}{\mathcal{B}}
\newcommand{\sC}{\mathcal{C}}
\newcommand{\sAB}{\sA{\cup}\sB}
\newcommand{\sigp}{\tilde{\sigma}^+}
\newcommand{\sigm}{\tilde{\sigma}^-}
\newcommand{\sigz}{\tilde{\sigma}^z}
\newcommand{\sigx}{\tilde{\sigma}^x}
\newcommand{\sigy}{\tilde{\sigma}^y}

\begin{document}
\title{Systematic construction of quantum many-body scars in frustrated Rydberg arrays}
\author{Jean-Yves Desaules}
\affiliation{Institute of Science and Technology Austria (ISTA), Am Campus 1, 3400 Klosterneuburg, Austria}
\author{Aron Kerschbaumer}
\affiliation{Institute of Science and Technology Austria (ISTA), Am Campus 1, 3400 Klosterneuburg, Austria}
\author{Marko Ljubotina}
\affiliation{Physics Department, Technical University of Munich, TUM School of Natural Sciences, Lichtenbergstr. 4,
Garching 85748, Germany}
\affiliation{Munich Center for Quantum Science and Technology (MCQST), Schellingstr. 4, M\"unchen 80799, Germany}
\author{Maksym Serbyn}
\affiliation{Institute of Science and Technology Austria (ISTA), Am Campus 1, 3400 Klosterneuburg, Austria}
\date{\today}

\begin{abstract}
Quantum many-body scars in Rydberg atom arrays have thus far only been observed on bipartite lattices, leaving open the question of whether and how they survive frustration, and what the appropriate initial states are that lead to nonthermal dynamics. We introduce a graph-theoretic framework to find suitable candidates for scarring on arbitrary lattices. Our framework predicts two distinct mechanisms: type-I scars generalize the bipartite case by using locally entangled states to overcome mild frustration, while type-II scars exploit strong frustration to pin part of the lattice, leaving the remainder to oscillate freely. We numerically demonstrate both mechanisms and uncover an exponential family of scarred trajectories on the hexagonal lattice that can encode information protected from thermalization. Our results establish scarring as a generic feature of Rydberg systems beyond one dimension and provide an experimentally accessible route to systematically probing non-thermal dynamics in quantum simulators.
\end{abstract}

\maketitle

\noindent{\bf \em Introduction.---}
Typically, unitary dynamics in chaotic interacting quantum systems lead to rapid transfer of information to non-local degrees of freedom, making the recovery of short-range-entangled (SRE) initial states near impossible. One way of avoiding this scrambling is to have an extensive number of conserved charges such that all the information is contained in them, as is the case in integrable~\cite{sutherland2004beautiful} and many-body localized systems~\cite{Abanin2019Colloquium, HuseReview}. 
Recently, a more subtle violation of thermalization was discovered in the form of quantum many-body scars (QMBSs)~\cite{Moudgalya2022Review, Serbyn2021Review, Chandran2023Review, Papic2022}. This phenomenon is characterized by the presence of a small number of non-thermal eigenstates living in a special subspace endowed with additional symmetries. If there exists an SRE state supported predominantly by these QMBSs, its dynamics may avoid thermalization.

QMBSs were first observed in an experiment in a one-dimensional (1D) Rydberg atom array~\cite{Bernien2017Rydberg}, approximately described by the PXP model~\cite{FendleySachdev, Lesanovsky2012} introduced below. Similar QMBSs have since been found in a wide variety of 2D lattices~\cite{MichailidisPRX, Bluvstein2021Controlling, Lin20202D, Huan20212D}. In Ref.~\cite{Bluvstein2021Controlling}, a Rydberg atom platform was used to probe multiple lattices, including square, honeycomb, Lieb, and hexagonal. Crucially, all lattices in which scarring was observed were bipartite, and the initial state exhibiting non-ergodic behavior was equivalent to a Néel state, with atoms in one sublattice in the Rydberg state and atoms in the other sublattice in the ground state. While non-bipartite lattices were also investigated in Ref.~\cite{Bluvstein2021Controlling}, no signature of scarring was found in these cases. 

The existence of scars in non-bipartite lattices is the central question our work addresses. 
Finding new scars beyond 1D is challenging, as most of the established methods and tools rely on matrix product states~\cite{ren2025scarfinder, petrova2025tdvp}, necessitate a known scar state as input~\cite{Szoldra2022QVAE}, or are limited to scarring from Fock states~\cite{Cao_2024}. Our work introduces a graph-theoretic approach that is dimension-insensitive and is free of these limitations, as it relies on the algebraic structure underlying QMBSs.

Specifically, our framework determines the best candidate for scarring in arbitrary lattices and the observables showing the largest oscillation amplitude. It leads to the identification of two distinct types of scarring, imposing different conditions on the underlying lattice, denoted as type-I/II. Type-I is a generalization of the bipartite case to appropriate non-bipartite lattices, which relies on short-range entanglement to overcome the frustration of the lattice. Type-II is distinct from previously observed scarring, demonstrating that 
frustration, conventionally an obstacle to coherent dynamics, can \emph{stabilize} a new class of scars. We note that, as type-I scars do not rely on frustration, their construction is not limited to non-bipartite lattices. Using our framework, we uncover an exponential number of scarred initial states in the hexagonal lattice. We confirm the results of our approach via numerical simulations, demonstrating the existence of QMBSs in the PXP model on multiple lattices and opening the door to their experimental studies. 

\noindent{\bf \em PXP model.---}
Rydberg atoms act as two-level systems, and in the blockade regime, neighboring atoms cannot be excited simultaneously. As a consequence, the system can be well described by the effective spin-1/2 PXP~\cite{FendleySachdev, Lesanovsky2012} Hamiltonian  
\begin{equation}\label{eq:PXP}
    H{=}\sum_{j}\tilde\sigma^x_j, \quad \tilde \sigma^x_j = \sigma^x_j \mathcal{P}_j, 
    \quad \mathcal{P}_j=\hspace{-0.25cm}\prod_{k: \langle j,k\rangle}\hspace{-0.27cm}P_k=\hspace{-0.25cm}\prod_{k: \langle j,k\rangle}\hspace{-0.27cm}\ket{\downarrow}\bra{\downarrow}_k,
\end{equation}
where the projectors $\mathcal{P}_j$ enforce the Rydberg blockade by allowing a spin to flip only if all of its nearest-neighbors (denoted as $\langle \cdot, \cdot \rangle$) are pointing down. The operators $\tilde \sigma^\alpha_j$ denote Pauli matrices (with $\alpha{=}x, \, y,\, z$) dressed by $\mathcal{P}_j$. 

In bipartite lattices, quenching from a N\'eel-like state leads to coherent oscillations instead of thermalization. This periodic motion has the wave-function transferring back and forth between the initial state and the other N\'eel state~\cite{Bernien2017Rydberg, TurnerNature, Bluvstein2021Controlling}. This dynamics is akin to the precession of a big spin and can be linked to an approximate SU(2) structure that is present only in the scarred subspace~\cite{Choi2018su2}. The two N\'eel states are respectively the lowest- and highest-weight states of $J^z$, while the Hamiltonian is proportional to the $J^x$ operator, leading to precession in the YZ plane.

\noindent{\bf \em Graph formalism and su(2) algebra.---}
In order to investigate arbitrary lattices, we will represent them as graphs $G(V,E)$. The vertices $v{\in} V$ are sites where atoms reside, and two vertices are connected by an edge $e{\in} E$ if they are in blockade range of each other. The Hamiltonian is still the one in Eq.~\eqref{eq:PXP}, with nearest-neighbors $\langle \cdot, \cdot \rangle$ now meaning vertices linked by an edge.

Our construction of scarred initial states will rely on separating all vertices into three disjoint sets  $\sA$, $\sB$, and $\sC$ such that $\sA\cap\sB{=}\sA\cap \sC{=} \sB\cap \sC{=}\emptyset$ and $\sA\cup \sB \cup \sC{=}V$. 
Sets $\sA$ and $\sB$ act as the two main sublattices between which occupation switches, while $\sC$ is a ``dead'' sublattice and its dynamics is strongly suppressed. We note that $\sC$ can be empty (as is the case for scarring in bipartite lattices) while $\sA$ and $\sB$ cannot, and that the latter two usually have the same cardinality up to boundary terms.

From this separation of the vertices, we build the approximate su(2) algebra linked to scarring from the raising and lowering operators~\footnote{We note that $J^+$ as defined here is only a proper raising operator if $\sC=\emptyset$, as otherwise it is not nilpotent due to the presence of $\sigx$.} 
\begin{equation}
	J^+=\sum_{j\in \sA}\sigp_j {+}\sum_{j\in \sB}\sigm_j {+}\frac{1}{2}\sum_{j\in \sC}\sigx_j \ \text{and} \ J^-=\left(J^+\right)^\dagger.
\end{equation} This allows us to define the $J^y$ operator as 
\begin{equation}
	J^y=i(J^-{-}J^+)/2=\sum\nolimits_{j\in \sA}\sigy_j -\sum\nolimits_{j\in \sB}\sigy_j, \label{eq:Sy}
\end{equation}
while the $J^z$ operator is $J^z=\frac{1}{2}\left[J^+,J^-\right]$, yielding
\begin{equation}
	J^z{=}\frac{1}{2}\sum_{j\in \sA}\hspace{-0.08cm}\left[\sigz_j{+}\hspace{-0.25cm}\sum_{k: \langle j, k\rangle}\hspace{-0.23cm}\left(\sigp_j\sigm_k{+}\sigp_k\sigm_j\right)\epsilon^{\sA}_k\right]\hspace{-0.08cm}{-}\left(\sA{\leftrightarrow} \sB \right). \label{eq:Sz}
\end{equation}
Here $\epsilon^{\sA}_k$ ($\epsilon^{\sB}_k$) is $1$ if $k{\in} \sA$ ($\sB$), 0.5 if $k {\in} \sC$ and 0 if $k\in \sB$ ($\sA$). 
Finally, we also have $J^x{=}\left(J^+{+}J^-\right)/2 {=}H/2$. 
If the algebra is closed, the ground state $\ket{{\rm GS}(\theta)}$ of the operator $J(\theta){=}\cos(\theta)J^z{-}\sin(\theta)J^y$ will evolve as $\ket{{\rm GS}(\theta{+}t)}$. If it is only approximate, the revivals will decay over time, as in the bipartite case~\footnote{As the system is initialized in the lowest-weight state of $J^z$ (up to a global rotation). This approximate closure only needs to hold for the SU(2) representation with maximum total spin.}.

\begin{figure}[t]
	\centering
	\includegraphics[width=\columnwidth]{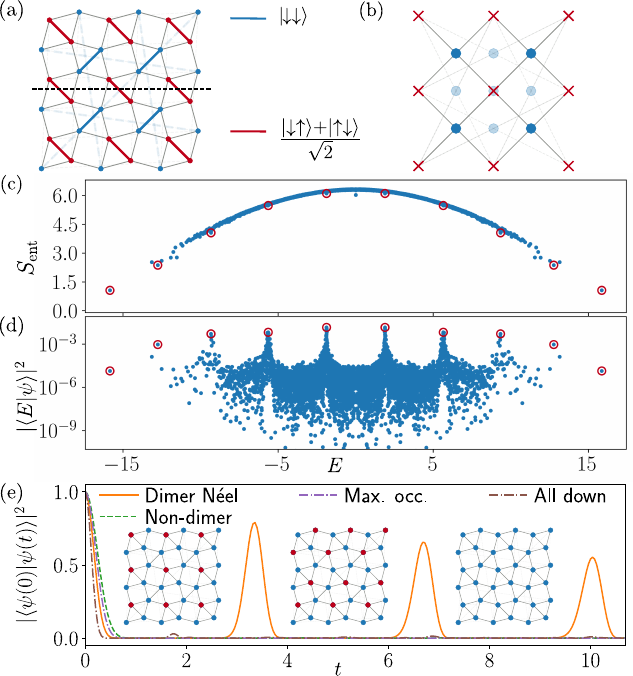}
	\caption{Type-I scarring in the Shastry-Sutherland lattice on a torus with 36 spins. (a) Dimer cover and (b) effective square lattice of these dimers. Atoms in $\sA$ ($\sB$) are in blue (red). (c) Entanglement entropy of eigenstates in the fully symmetric sector along the cut shown in (a). (d) Overlap of the dimer N\'eel state shown in (a) with the eigenstates in the fully symmetric sector. States highlighted in red in (c) and (d) are the same and are those with maximal overlap in their energy range. While the scarred structure is clearly visible in panel (d), there is no band of well-separated eigenstates. As a consequence, no clear outliers are visible in the entanglement entropy, making it hard to find scars using this data alone. (e) Return fidelity over time for various initial states. While the structure of the dimer N\'eel state is shown in (a),  the corresponding $\{\sA,\sB\}$ partitions for the three other states are shown below them. Only the dimer N\'eel state displays revivals.
	}
	\label{fig:ShaSu}
\end{figure}

Through this algebraic construction, we have reduced the task of finding an initial state to that of finding a suitable partition of $V$ into $\sA$, $\sB$, and $\sC$. Each trial $\{\sA,\sB,\sC\}$ can be tested by quenching from the ground state of $J^z$ or $J^y$ and monitoring these same two observables. We emphasize that, as both are sums of local terms, the initial states will be SRE, allowing for efficient preparation via annealing or adiabatic ramping from a product state. We now propose two different rules for finding partitions $\{\sA,\sB,\sC\}$ that lead to scarring. 

\noindent{\bf \em Type-I scarring.---}
For type-I scarring, the idea is to map the physical, non-bipartite lattice to an effective bipartite lattice. Let $S=\{s_j\}_j$ be a set of subsets $s_j$ of vertices which are disjoint and cover all vertices of $G$. We postulate that type-I scarring exists if: 
\begin{enumerate}[label=(\roman*).]
    \item{$S$ is a clique cover: the induced subgraph $G[s_j]$ is complete for any $s_j$.}
    \item{The quotient graph $G_S{=}G/S$ is bipartite.}
\end{enumerate}
The first condition implies that all atoms in any subset $s_j$ are blocking each other, ensuring that it can have at most one up-spin. Hence, the symmetric subspace in each $s_j$ only contains the all-down and $W$ states, with the Hamiltonian in Eq.~\eqref{eq:PXP} flipping between them. This allows us to consider the $s_j$ as spin-1/2 degrees of freedom of an approximate model on the lattice $G_S$. Meanwhile, the second condition means that $G_S$ is bipartite, allowing us to divide all $s_j$ into $\sA$ and $\sB$ such that the effective sites in each group do not block each other. There are thus two N\'eel states to oscillate between in this effective lattice, and they take the following form in the physical lattice $G$: all $s_j$ in sublattice $\sA$ ($\sB$) are in the all-down configuration, while all $s_j$ in sublattice $\sB$ ($\sA$) are in the $W$-state. These states are respectively the ground and ceiling states of $J^z$ as defined in Eq.~\eqref{eq:Sz}. Indeed, the $\pm\sigz$ promote down-spins in $\sA$ ($\sB$) and up-spins in $\sB$ ($\sA$) while the $\sigp\sigm$ delocalize these up-spins across all sites in each $s_j$. 

Using this construction, we unveil scarring in the Shastry-Sutherland (or square-triangular) lattice. It is shown on Fig.~\ref{fig:ShaSu} (a), along with the only clique cover satisfying all requirements for type-I scarring. In this figure and in the rest of this work, we will use the convention that atoms in $\sA$ are in blue, atoms in $\sB$ in red, and atoms in $\sC$ in black. The effective lattice $G_S$ is a square lattice, shown in panel (b). While no strong outliers are visible in the entanglement entropy $S_\mathrm{ent}$ of eigenstates, the hidden scarred structure is revealed when taking their overlap with the dimer N\'eel state, see panel (d). Consequently, strong revivals are seen in the return fidelity $|\langle \psi(0)|\psi(t)\rangle|^2$. This quantity measures how much of the wave function returns to its initial configuration over time and is shown on panel (e).

\begin{figure}[t]
	\centering
	\includegraphics[width=\columnwidth]{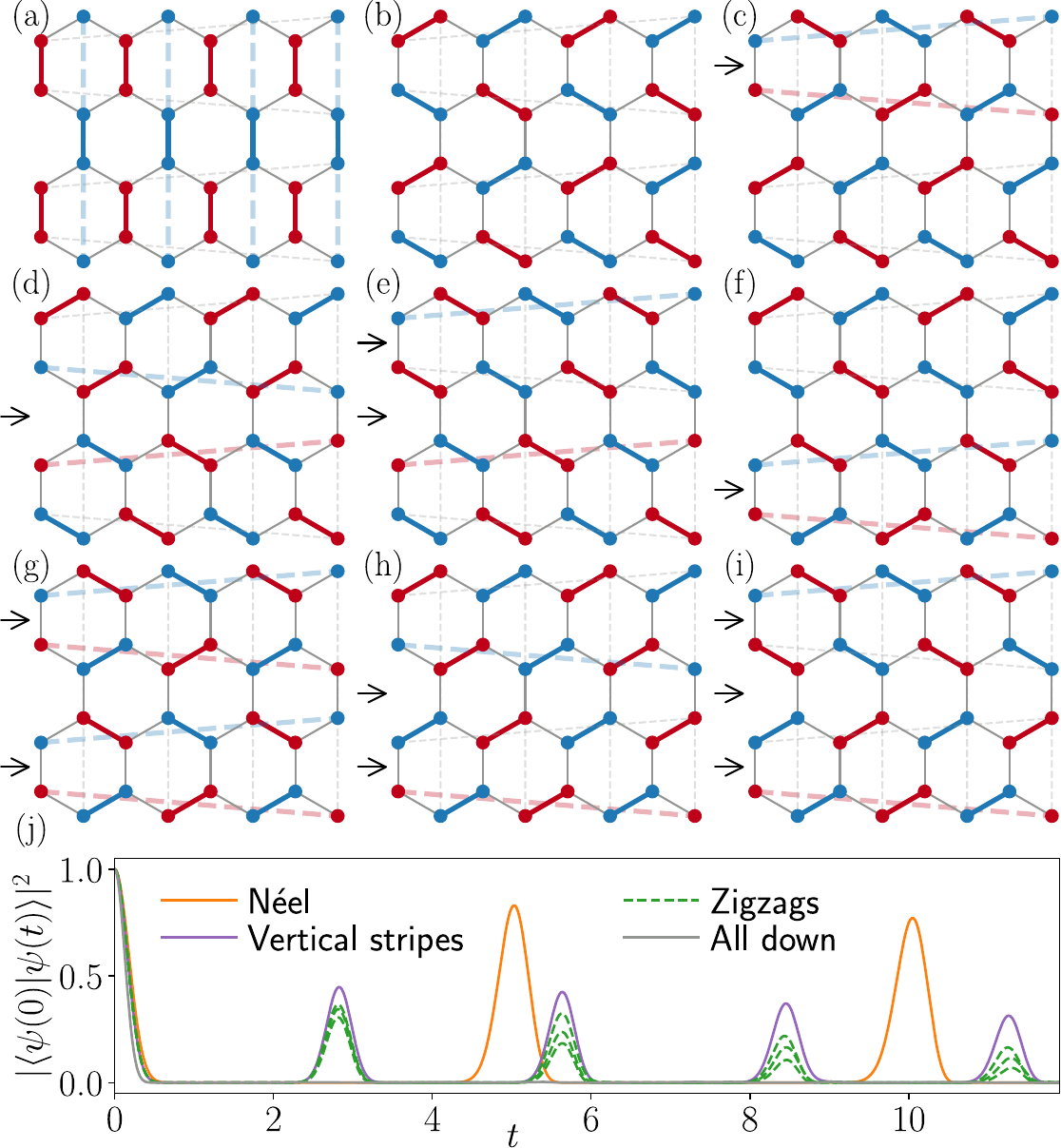}
	\caption{Type-I scarring in the hexagonal lattice on a torus with $N_x{=}4$, $N_y{=}4$ and 32 spins. (a)-(i) The 9 bipartite (out of 449) dimer covers. (a) Vertical stripe cover. (b)-(i) The $2^{N_y-1}{=}8$ different zigzag dimer covers. Black arrows indicate which rows are shifted. (j) Fidelity after quenches from the initial states shown in (a)-(i), the N\'eel and all down states.
	}
	\label{fig:Hex}
\end{figure}

We note that our construction also encompasses known cases of scarring in various lattices.
If $G$ is bipartite, we choose each $s_j$ to contain a single vertex, and the quotient graph is equal to $G$ itself. This covers scarring from the N\'eel state in the 1D PXP model as well as in the 2D bipartite lattices as explored in Ref.~\cite{Bluvstein2021Controlling}. 
Another known case is when atoms in each subset $s_j$ have a permutation symmetry. Taking the symmetric subspace of a subset then \emph{exactly} maps to a two-level system. In Rydberg atoms, these types of graphs have been studied in relation to the maximum independent set problem in Refs.~\cite{schiffer2024circumventing, lukin2024quantum}. We note that, in those works, the $s_j$ in $\sA$ and $\sB$ have different cardinalities, with sets in $\sA$ containing a single vertex and those in $\sB$ containing two vertices. This is not forbidden by our construction but usually leads to worse scarring signatures (see the discussion of the ruby lattice in the End Matter).
Finally, 1D Rydberg arrays with longer blockade ranges~\cite{kerschbaumer_quantum_2024} can be recast as PXP models on ladders, and their QMBSs with even spatial periodicity also follow the type-I construction.

\noindent{\bf \em Systematic search for type-I partitions.---}
While it is straightforward to check that these known examples of scarring fit into our construction, one can wonder about the practicality of finding $s_j$ that satisfy conditions (i) and (ii) for an arbitrary graph $G$. First, given $S$ it is easy to build $G_S$ and to check if it is bipartite, so the crux of the problem is finding $S$ such that (i) is satisfied. In fact, finding all such $S$ is equivalent to finding all clique covers of $G$. While this is NP-hard in general, we argue that it is still tractable in the highly structured lattices used in quantum systems. Indeed, the largest cliques are small, usually being at most 3-cliques (triangles) in 2D and 4-cliques (tetrahedra) in 3D. One can then use Knuth's dancing links algorithm (DLX)~\cite{knuth2000dancinglinks} to find all exact covers spanned by these cliques. This can be sped up significantly if one restricts to cliques of the same size (which typically lead to better scarring). While this implies running the DLX algorithm for each clique size, these instances are much faster due to the smaller number of cliques.

This approach allows one to find \emph{all} possible covers that satisfy the conditions for type-I scarring. For example, in the (non-bipartite) ruby lattice, we find two different scarred initial states (see End Matter). This method also works in bipartite cases, and in the hexagonal lattice, our approach reveals an \emph{exponential} number of scarred states. While the exact number depends on the boundary conditions, we focus on the lattice on a torus with $N_x$ hexagons in one direction and $N_y$ in the other. An example is shown in Fig.~\ref{fig:Hex} for $N_x{=}N_y{=}4$. While for any $N_x$ we find a dimer cover with stripe order as shown in panel (a), if $N_x$ is even, we find a total of $2^{N_y-1}$ dimer covers in zigzag patterns, see panels (b)-(i). This exponential scaling with $N_y$ arises because conditions (i) and (ii) remain unaffected when the dimers are shifted by one site along two horizontal zigzags, as shown for the first two in panel (b). As such, a shift can be done along the lines around each of the $N_y$ rows of hexagons. 
Since shifting all rows returns the original cover (up to an irrelevant swap of $\sA$ and $\sB$), there are only $N_y-1$ independent shifts, leading to $2^{N_y-1}$ different dimer covers.
Thanks to the long coherence time of scars, this allows the preservation of $N_y-1$ bits of information from thermalization.

While the type-I scarring construction is powerful, it does not allow for finding scars in all non-bipartite lattices. We did not find any $S$ that would satisfy the criteria in the triangular and kagome lattices. More generally, lattices composed purely of triangles seem particularly resistant to type-I scarring due to their high connectivity and large frustration. However, the same features can also be used to promote a different construction, which we call type-II scarring.

\noindent{\bf \em Type-II scarring.---}
Unlike in type-I discussed previously, we now consider a non-empty sublattice $\sC$. This third sublattice will act as a buffer to prevent sublattices $\sA$ and $\sB$ from blocking each other. At the same time, $\sC$ should not interfere with the oscillations in $\sA$ and $\sB$, and its dynamics should be strongly suppressed. As such, we postulate that type-II scarring occurs when~\footnote{We note that these conditions are not strictly necessary. As in type-I scarring, one could theoretically consider cliques instead of single vertices in $\sA$ and $\sB$ and some level of direct interactions between these two sublattices. However, in practice, we did not find any such case where scarring was prominent.}
\begin{enumerate}[label=(\Roman*).]
	\item{All vertices in $\sA$ and $\sB$ only have neighbors in $\sC$.}
	\item{Each vertex in $\sC$ is connected to at least one vertex in $\sA$ and one vertex in $\sB$.}
	\item{The induced subgraph $G[\sC]$ is strongly connected, or the number of neighbors in $\sAB$ that each vertex in $\sC$ has is greater or equal to the number of neighbors in $\sC$ that each vertex in $\sAB$ has.}
\end{enumerate}
Condition (I) ensures that $\sA$ and $\sB$ do not blockade each other, while conditions (II) and (III) are there to make sure that excitations in $\sC$ are suppressed. While the latter two might seem redundant, we illustrate the need for condition (III) in the End Matter.

One advantage of type-II scarring over type-I is that, due to (I), all terms in the $J^y$ operators in Eq.~\eqref{eq:Sy} are non-overlapping and thus commute with one another. Indeed, $\sigma^y$ terms act only on sites in $\sAB$ while the projectors act on sites in $\sC$. Additionally, each site in $\sC$ will have at least one projector applied to it due to condition (II). As such, the ground state $\ket{\mathrm{GS}_Y}$ of this operator will have all sites in $\sC$ down and the sites in $\sA$ and $\sB$ in the $\pm 1$ eigenstate of $\sigma^y$. Hence $\ket{\mathrm{GS}_Y}$ is a product state and can be prepared in Rydberg atom arrays with a single site-dependent pulse along X. Meanwhile, the ground state of $J^z$ generally has some amount of entanglement due to the competition between the $\sigz$ and $\sigm\sigp$ terms. We will focus on $\ket{\mathrm{GS}_Y}$ as the preferred initial state. In the same spirit of experimental relevance, we will show the occupation $n_j=(1+\sigma^z_j)/2$ of sites in different sublattices after a quench. This quantity is directly measurable in Rydberg atom platforms and should exhibit clear oscillations due to its large overlap with $J^z$.

\begin{figure}[t]
	\centering
	\includegraphics[width=\columnwidth]{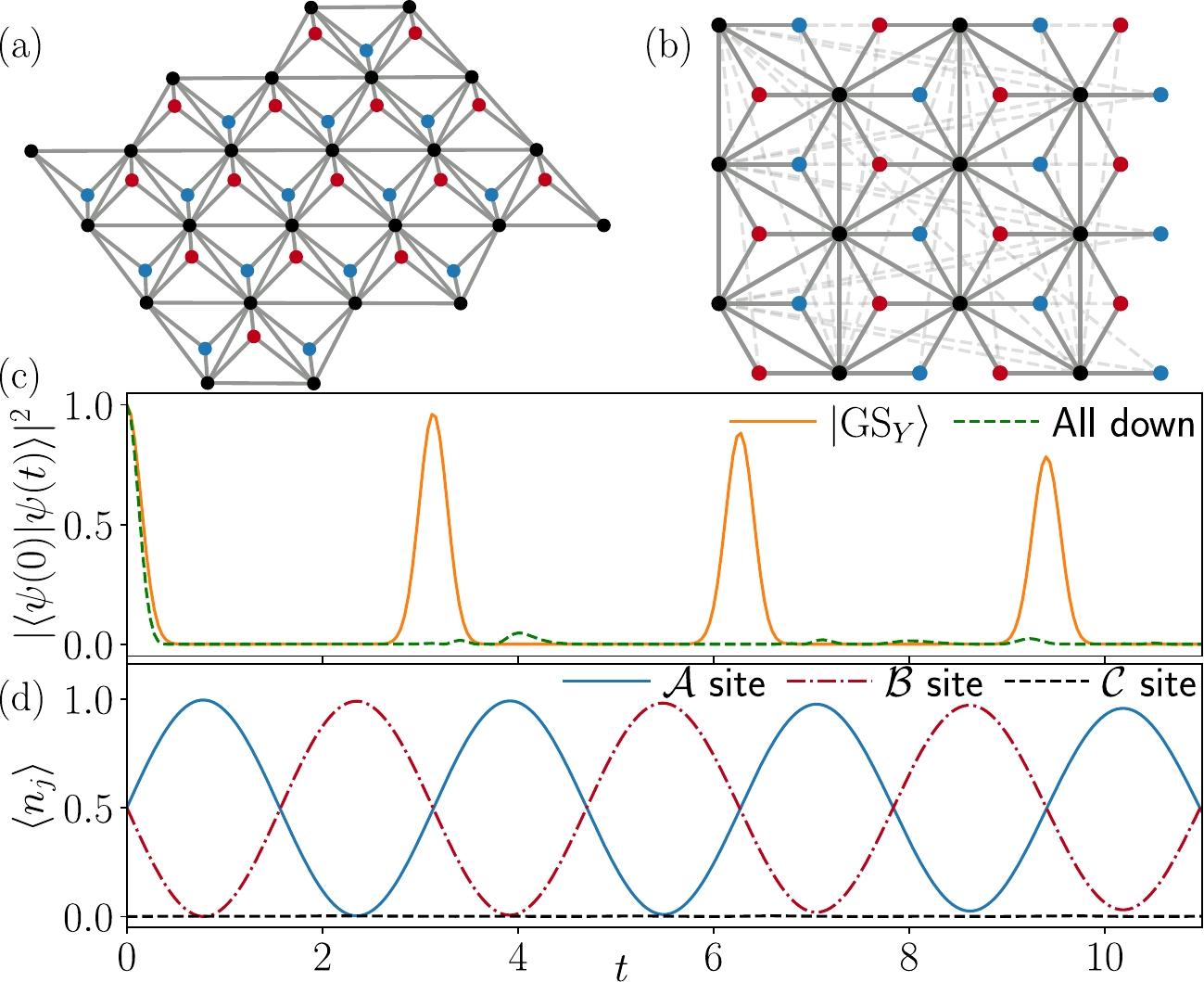}
	\caption{Type-II scarring in the quasi-2D asanoha lattice. (a) The lattice is a triangular lattice made of tetrahedra that point up and down alternatively. (b) Lattice with 36 spins used for the numerical simulations in (c) and (d). (c) Fidelity after a quench from two different initial product states. (d) Occupation $n_j=(1+\sigma^z_j)/2$ after a quench from the ground state of $J^y$. The occupation of sites in $\sC$ is highly suppressed, while $\sA$ and $\sB$ oscillate almost freely. 
	}
	\label{fig:FC_tri}
\end{figure}

In the previous section, we discussed the absence of type-I scarring in the triangular and kagome lattices. While these lattices also do not show strong type-II scarring (see End Matter), their quasi-2D counterparts do. These lattices are now composed of tetrahedra instead of triangles, further increasing the frustration. This makes it possible to satisfy condition (III). 
In Fig.~\ref{fig:FC_tri}, we show clear signatures of scarring for a quasi-2D asanoha lattice. The suppression of dynamics in $\sC$ is quite visible in panel (d), allowing $\sA$ and $\sB$ to oscillate almost perfectly. The same approach can be used in the kagome lattice to obtain near-perfect scarring in a quasi-2D pyrochlore lattice (see End Matter).

We note that in order to satisfy conditions (I) and (II), $\sAB$ has to be a maximal independent set. Meanwhile, condition (III) favors the largest possible $\sAB$, meaning that the best candidates are maximum independent sets. One can rely on standard tools for finding such sets, including Rydberg atom simulators~\cite{pichler2018MIS, Ebadi2022MIS}. Finding their optimal partition into $\sA$ and $\sB$ can be solved using discrete optimization techniques, with a cost function that incorporates measures of connectivity between $\sA$ and $\sC$ and between $\sB$ and $\sC$. 

\noindent{\bf \em Discussion.---}
In this Letter, we demonstrated that scarring in Rydberg atom arrays is common, even in non-bipartite lattices.
Our work also illustrates that the relation between frustration and QMBSs can be of two different kinds. In type-I scarring, a small amount of frustration can be overcome by using locally entangled initial states. This type of scarring is similar to that observed previously in bipartite lattices. Meanwhile, if the frustration is strong enough, it can pin down part of the lattice and suppress its dynamics, allowing the rest of the system to oscillate almost freely. We call this novel behavior in constrained systems type-II scarring. 

The only cases in which we do not find strong scarring are in the triangular and kagome lattices, where frustration is too large for type-I but too low for type-II, raising the question of whether these models still may feature nonthermal dynamics. Other open questions concern the robustness of our scars away from the ideal PXP limit and the potential coexistence of type-I and type-II scars. Finally, as in bipartite lattices, both $\sA$ and $\sB$ are maximum independent sets (MISs) while in type-II scars $\sAB$ is also an MIS. Our work deepens the connection between such sets and QMBSs~\cite{lukin2024quantum} and invites the study of potential links in other constrained models.

Scarred oscillations in the 1D PXP model were also recently shown to be a suitable vacuum for coherent soliton-like excitations~\cite{kerschbaumer_discrete_2025}, possibly being connected to superdiffusive transport~\cite{LjubotinaPRX, morettini_2025}. It is therefore natural to ask whether the type-I/II scarred states uncovered here could similarly act as a vacuum for more complex non-thermal dynamics and lead to anomalous energy transport in these systems.

While our numerical results are limited to modest system sizes (with still large Hilbert spaces), we argue that testing our predictions in larger arrays is best suited to quantum simulators. In fact, making the search for scarring in these platforms more efficient is precisely one of the main contributions of our work. Using the sets of rules we have laid down, rapid testing for the presence of scarring in a large number of lattices can be performed in modern Rydberg atom arrays. Such platforms are moreover well placed to probe non-thermal physics beyond clean revivals: a recent experiment on the 1D PXP model~\cite{mark2025observationballisticplasmamemory} showed that even quenches with no revivals can host ballistic plasma quasiparticles and long-lived memory of charge clusters, exposed by precision diagnostics.
Similar observables may likewise reveal more subtle non-thermal signatures potentially appearing at long times, also in cases where scarring signatures are short lived such as in the triangular and kagome lattices.

\textit{Note added.}—During completion of this work, we became aware of a complementary study that unveils quantum many-body scars in a frustrated Rydberg atom array, stemming from the structure of the Hamiltonian adjacency graph~\cite{Verde2026geometry}.

\begin{acknowledgments}
\noindent{\bf \em Acknowledgments.---}
We acknowledge useful discussions with Zlatko Papi\' c, Ana Hudomal, and Jie Ren. J.-Y.D.~acknowledges funding from the European Union's Horizon 2020 research and innovation programme under the Marie Sk\l odowska-Curie Grant Agreement No.~101034413.
M. L. acknowledges support by the Deutsche Forschungsgemeinschaft (DFG, German Research Foundation) under Germany’s Excellence Strategy – EXC-2111 – 390814868.
This research was funded in whole or in part by the Austrian Science Fund (FWF) [10.55776/COE1]. For Open Access purposes, the author has applied a CC BY public copyright license to any author-accepted manuscript version arising from this submission. We also acknowledge support by the Erwin Schr\"odinger International Institute for Mathematics and
Physics (ESI).
\end{acknowledgments}

\bibliography{biblio}

\clearpage 
\pagebreak
\begin{appendix}
	\begin{center}
		{\bf \large End Matter}
	\end{center}

\begin{figure}[b]
	\centering
	\includegraphics[width=\columnwidth]{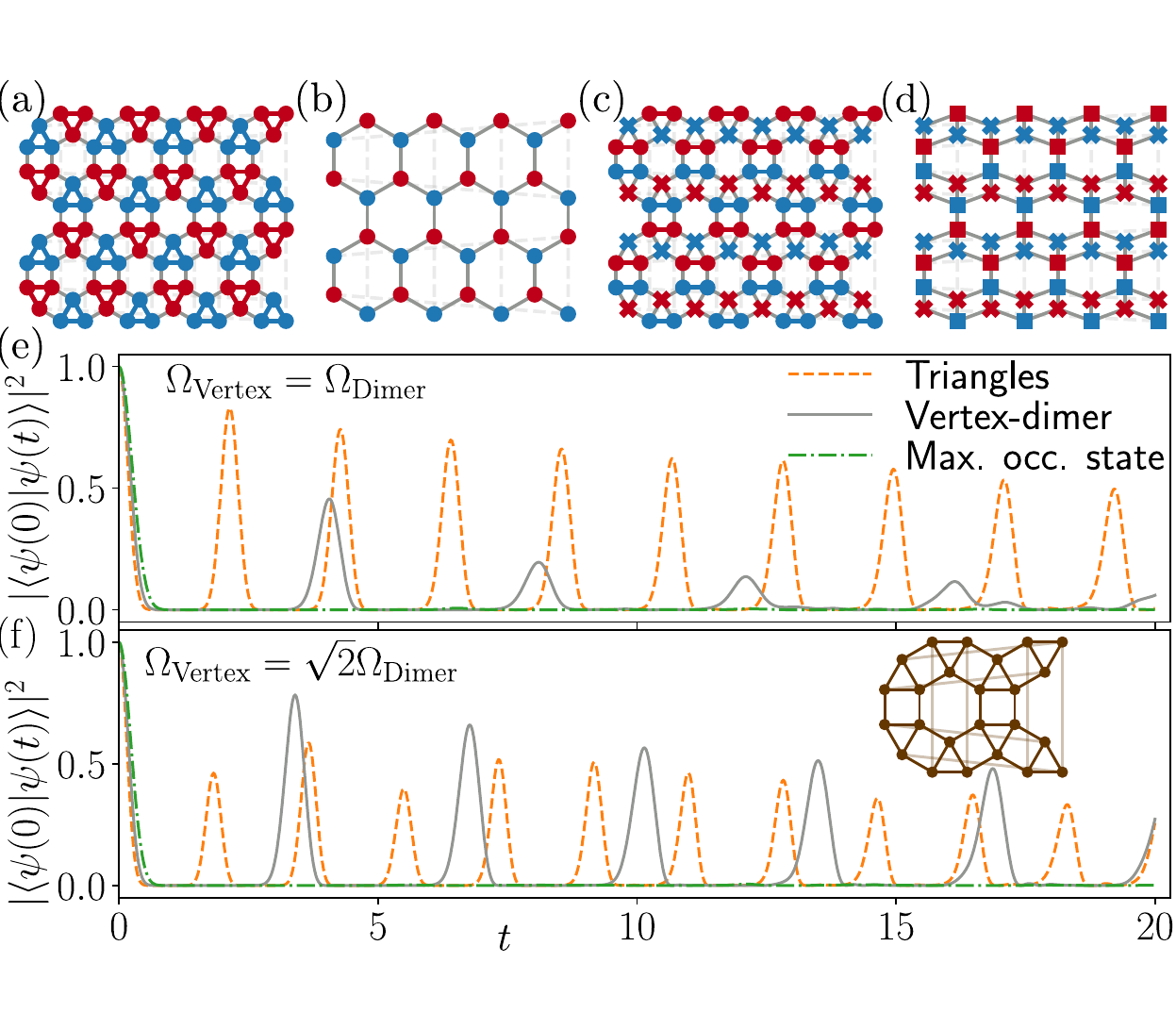}
	\caption{(a) Triangle cover in the ruby lattice and (b) corresponding effective hexagonal lattice. (c) Vertex-dimer cover and (d) corresponding effective lattice. The squares in (d) correspond to dimers, while crosses correspond to single vertices in the original lattice. (e)-(f) Fidelity after quenches from various initial states in a system with 24 sites, see inset of (f). The maximally occupied (Max. occ.) state is one of the basis states with the most up-spins. With a homogeneous Hamiltonian, the best revivals are seen from the N\'eel state of the triangle cover, while in the frequency-adjusted Hamiltonian, the vertex-dimer N\'eel state displays the best fidelity instead. 
	}
	\label{fig:Ruby}
\end{figure}

\noindent{\em \bf  Ruby lattice.---}
In this section, we show that for the non-bipartite ruby lattice, there are two different clique covers that satisfy the conditions for type-I scarring. These are shown in Fig.~\ref{fig:Ruby} along with the effective bipartite lattice. The first cover is composed of triangles, while the second is a mix of dimers and single vertices. While both show scarring, when evolving the system with the Hamiltonian in Eq.~\eqref{eq:PXP}, the return fidelity for the vertex-dimer N\'eel state decays rapidly. This is due to the effective frequency being different for dimers and single vertices. Indeed, if an effective two-level system is composed of $k$ sites, the effective flip term between the all-down state and the $W$ state will be proportional to $\sqrt{k}$. So if subsets $s$ have different cardinalities, they will precess with different frequencies. Over time, neighboring subsets will become excited at the same time, leading to blockaded interactions and the decay of oscillations.

To stabilize scarring in such a case, one can consider using a non-homogeneous Hamiltonian $H{=}\sum_{j}\Omega_j\sigx_j$. By adjusting the driving frequency of the atoms in each set $s$ by a factor of $1/\sqrt{|s|}$, one recovers a homogeneous effective Hamiltonian. The effect of this operation can be seen in Fig.~\ref{fig:Ruby} (f) for the ruby lattice, leading to stronger revivals from the vertex-dimer N\'eel state.

\begin{figure}[t]
	\centering
	\includegraphics[width=\columnwidth]{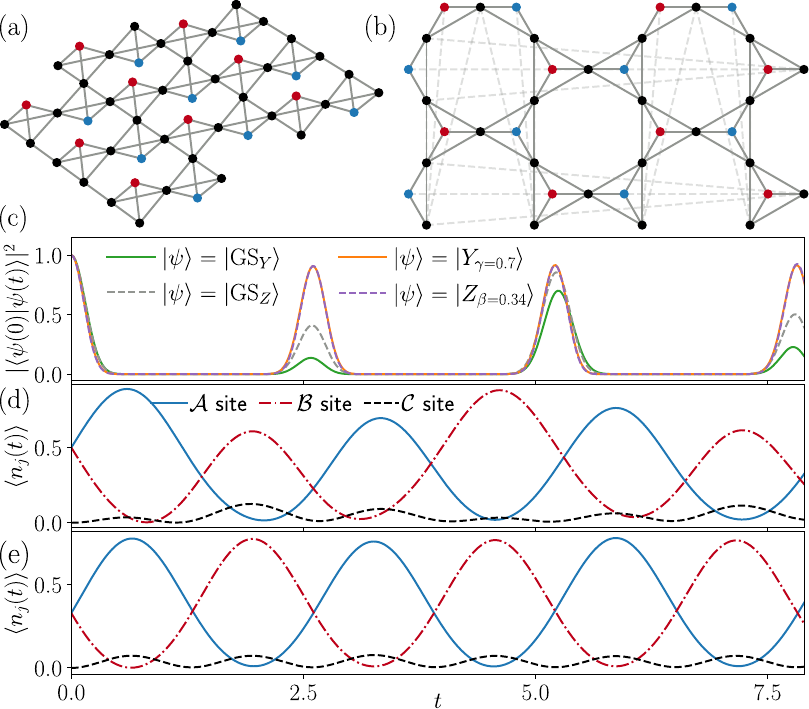}
	\caption{Type-II scarring in the quasi-2D pyrochlore lattice. (a) The lattice is a kagome lattice of tetrahedra, with tetrahedra pointing up and down alternately. (b) Lattice with 40 spins used for the numerical simulations in (c) and (d). (c) Return fidelity after quenches from $\ket{\mathrm{GS}_Z}$ and $\ket{\mathrm{GS}_Y}$ and from the deformed states. (d)-(e) Sublattice occupation after a quench from (d) $\ket{\mathrm{GS}_Y}$ and (e) $\ket{Y_{\gamma=0.7}}$. The occupation of $\sC$ is suppressed, but still shows some oscillations.
	}
	\label{fig:pyrochlore_all}
\end{figure}

\begin{figure}[t]
	\centering
	\includegraphics[width=\columnwidth]{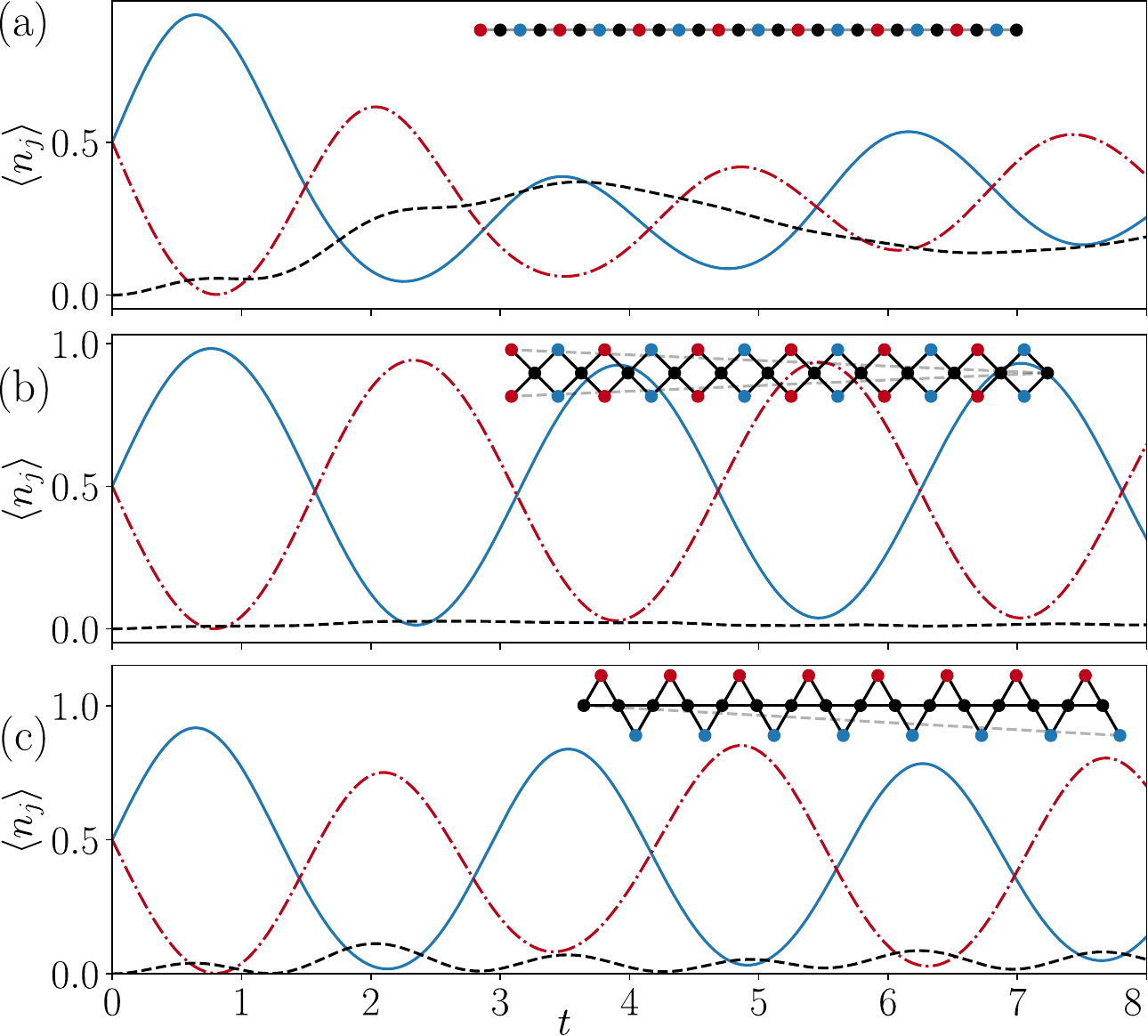}
	\caption{Signatures of type-II scarring in multiple quasi-1D lattices. (a) In the standard PXP model, there are still clear traces of type-II scarring from the $Y$ state with period 4. This can be made much more robust by using a PXP ladder as in (b), due to each $\sC$ site now being blockaded by two $\sA$ sites and two $\sB$ sites. Alternatively, connecting more sites in $\sC$ together as in (c) limits the number of excitations present in that sublattice, which also stabilizes scarring.
	}
	\label{fig:Type_II_1D}
\end{figure}

\noindent{\em \bf  Quasi-2D pyrochlore lattice.---}
In this section, we discuss type-II scarring in a quasi-2D pyrochlore lattice, as shown in Fig.~\ref{fig:pyrochlore_all}. As sites in $\sC$ only have two neighbors in $\sAB$ instead of six in the triangular case, occupation of $\sC$ is less strongly suppressed, and sites in $\sC$ still show some dynamics. Nonetheless, clear scarring can be seen from both $\ket{\mathrm{GS}_Y}$ and $\ket{\mathrm{GS}_Z}$. Almost perfect scarring can also be obtained in this model by slightly deforming the initial state. Let us define the state
\begin{equation}
	\ket{Y_\gamma}{=}\hspace{-0.1cm}\left(\!\!\bigotimes_{ \sA}\frac{\ket{\downarrow}{+}i\gamma\ket{\uparrow}}{\sqrt{1+|\gamma|^2}}\right)\!\!\otimes\!\!\left(\!\!\bigotimes_{ \sB}\frac{\ket{\downarrow}{-}i\gamma\ket{\uparrow}}{\sqrt{1+|\gamma|^2}}\right)\!\!\otimes\!\!\left(\!\!\bigotimes_{\sC}\ket{\downarrow}\right),
\end{equation}
for which we recover $\ket{Y_{\gamma=1}}=\ket{\mathrm{GS}_Y}$. We also define
\begin{equation}
	\ket{Z_\beta}{=}\frac{1}{\mathcal{N}}\left(\sum_{j\in \sB}\sigp_j+\beta\sum_{j \in \sC} \sigp_j\right)^{|\sB|}\ket{\downarrow \ldots \downarrow},
\end{equation}
with $\mathcal{N}$ a normalization factor. This state has $|\sB|$ excitations that can each be localized on a $\sB$ site or on one of the $\sC$ sites around it, while the $\sA$ are all down. Any excitation on a $\sC$ site is penalized by a factor of $\beta$. 
For small values of $\beta$, this is an SRE state where the excitations are mostly on $\sB$ sites but are slightly delocalized on neighboring $\sC$ sites. This mimics the structure of $\ket{\mathrm{GS}_Z}$, where the $\sigz$ terms favor having the excitation in $\sB$ while the $\sigp\sigm$ favor delocalisation. In fact, we find that $\ket{Z_{\beta=0.205}}$ has an overlap of 0.99 with $\ket{\mathrm{GS}_Z}$.

\begin{figure}[b]
	\centering
	\includegraphics[width=\columnwidth]{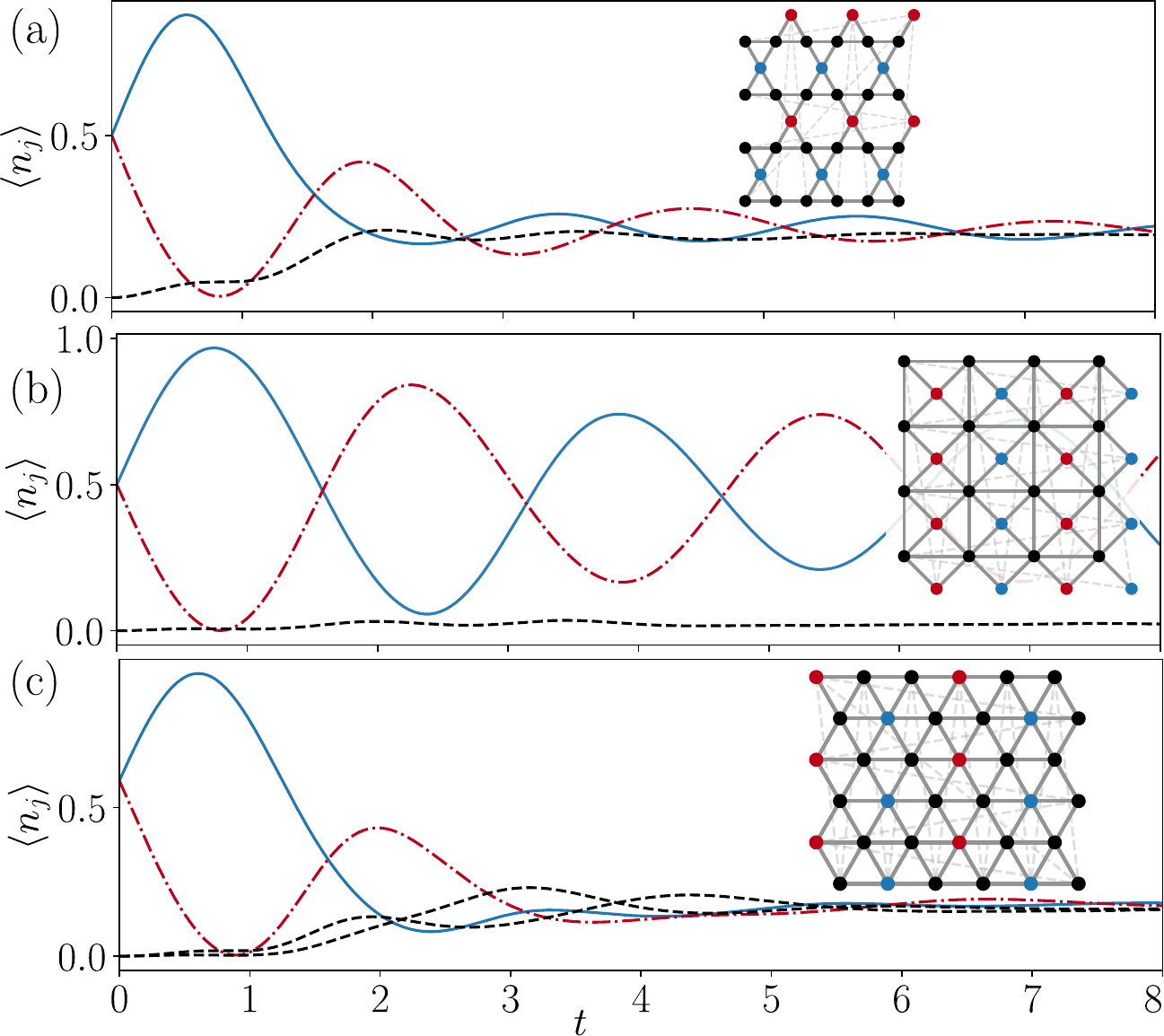}
	\caption{Signatures of type-II scarring in multiple 2D lattices (a) Despite being similar to the lattice in Fig.~\ref{fig:Type_II_1D} (c), in the kagome lattice, type-II scarring is not robust. (b) Clear scarring in the face-centered square lattice. (c) The triangular lattice shows dynamics similar to those of the kagome case, with oscillations only at short times.
	}
	\label{fig:Type_II_2D}
\end{figure}

Through numerical optimization based on the return fidelity, we find that the best scarring is obtained for $\gamma\approx 0.7$ and $\beta\approx 0.34$. These two states are located on the same orbit and display oscillations that show almost no decay in time, see Fig.~\ref{fig:pyrochlore_all} (c). We believe that these deviations from the ground states of the algebra operators are lattice-specific and cannot be predicted through our high-level graph approach.  

\noindent{\em \bf  Type-II scarring in other lattices---.}
Without condition (III), many lattices are seemingly compatible with type-II scarring, including the 1D chain and many other bipartite lattices. In this section, we show that satisfying only conditions (I) and (II) leads to unstable oscillations, and that modifying the lattice to ensure compliance with condition (III) vastly improves their lifetime. As in the main text, we will quench from $\ket{\mathrm{GS}_Y}$ and monitor the occupation on each sublattice. In Fig.~\ref{fig:Type_II_1D} (a), we show that in the 1D case, some coherent dynamics can be seen that are linked to type-II scarring.

These can be enhanced by either increasing the number of neighbors in $\sAB$ that each $\sC$ vertex has or by connecting the vertices in $\sC$ together more strongly. These two different approaches are shown in panels (b) and (c) of Fig.~\ref{fig:Type_II_1D}, respectively. In both cases, the occupation of sublattice $\sC$ is more strongly suppressed, leading to stable oscillations in the observables we consider. 

While the number of neighbors in $\sAB$ that each site in $\sC$ has is relevant, so is the number of neighbors in $\sC$ that atoms in $\sAB$ have. Indeed, expanding the model in Fig.~\ref{fig:Type_II_1D} (c) in the vertical direction yields the kagome lattice, as shown in Fig.~\ref{fig:Type_II_2D} (a). Using the same partition of vertices, one sees that the oscillations decay rapidly when compared to the quasi-1D case. This difference of behavior is likely due to sublattices $\sA$ and $\sB$ now having four neighbors in $\sC$ instead of two. $\sA$ and $\sB$ are thus more blockaded by $\sC$, preventing the build-up of excitations in them and leading to a faster decay of oscillations. 

In Fig.~\ref{fig:Type_II_2D} (b) and (c), we test type-II scarring in the face-centered square (FCS) and triangular lattices. These two cases also illustrate the importance of looking at both the number of neighbors in $\sAB$ that atoms in $\sC$ have and the number of neighbors in $\sC$ that atoms in $\sAB$ have. In the FCS case, both quantities are equal to 4 while in the triangular lattice the former is equal to 3 while the latter is equal to 6. This implies that condition (III) is satisfied in the FCS case but not in the triangular one. This matches well with the observed dynamics.

\end{appendix}

\end{document}